\documentclass[prd,showpacs,amsmath,showkeys,twocolumn,floatfix,amssymb, preprintnumbers, nofootinbib, superscriptaddress]{revtex4} 
\usepackage{epsfig,dcolumn}
\usepackage{graphicx}
\usepackage{comment} 
\usepackage[usenames]{color}
\usepackage{graphicx}
\usepackage{bm}
 \usepackage{ifpdf}

\begin{document}

\title{  A solvable three-body model in   finite volume }

\author{Peng~Guo}
\email{pguo@jlab.org}

\affiliation{Department of Physics and Engineering,  California State University, Bakersfield, CA 93311, USA.}

\author{Vladimir~Gasparian}
\affiliation{Department of Physics and Engineering,  California State University, Bakersfield, CA 93311, USA.}

\date{\today}

\begin{abstract}

In this work,   we propose an  approach to the solution of finite volume three-body problem by considering asymptotic forms and periodicity property of  wave function in configuration space. The asymptotic forms of wave function  define  on-shell physical transition amplitudes that  are  related to distinct dynamics, therefore,    secular equations of finite volume problem in this approach require only physical transition amplitudes. For  diffractive spherical part of wave function, it is  convenient to map a three-body problem into a higher dimensional two-body problem, thus, spherical part of solutions in finite volume resembles higher spatial dimensional two-body L\"uscher's formula.  The idea is demonstrated by an example of      two light spinless particles and one heavy particle scattering in one spatial dimension. 

  \end{abstract} 

\pacs{ }

\maketitle

{\it Introduction.}---Three-body interaction plays an important role in many aspects of nuclear, hadron and condensed matter physics. In nuclear and astrophysics for example, the precise knowledge of nucleon interaction is the key to understand nucleon structure and dynamics of nuclei, and it is also the fundamental information to explore the origin of universe. In hadron physics, three-body dynamics could be crucial in many physical processes, such as  extraction of   light quark mass difference  from  isospin violating decay of \mbox{$\eta \rightarrow 3 \pi$} \cite{Kambor:1995yc,Anisovich:1996tx,Colangelo:2009db,Lanz:2013ku,Schneider:2010hs,Kampf:2011wr,Guo:2015zqa,Guo:2016wsi,Colangelo:2016jmc}.  Three-body effect  also have  attracted a lot of interest in  condense matter physics, for examples, the fractional quantum Hall states \cite{Fradkin:1998,Cooper:2004} and cold polar molecules \cite{Buechler:2007}.   Traditionally, three-body dynamics has been studied based on many different approaches,  such as Bethe-Salpeter equations \cite{Taylor:1966zza,Basdevant:1966zzb,Gross:1982ny},   Faddeev's equation \cite{Faddeev:1960su,Faddeev:1965,Gloeckle:1983,Gloeckle:1995jg}, and dispersive approch \cite{Khuri:1960zz,Bronzan:1963xn,Aitchison:1965kt,Aitchison:1965zz,Guo:2014vya,Guo:2014mpp,Danilkin:2014cra,Guo:2015kla}.

Unlike the traditional three-body dynamics in free space, the finite volume three-body problem is still in a developing phase, though  some progresses from different approaches have been made in recent years, such as quantum field theory based diagrammatic approach and Faddeev equation  based methods  \cite{Kreuzer:2008bi,Kreuzer:2009jp,Kreuzer:2012sr,Polejaeva:2012ut,Briceno:2012rv,Hansen:2014eka,Hansen:2015zga,Hansen:2016fzj}, and the approach by considering periodic wave function in configuration space \cite{Guo:2016fgl}. In contrast, the finite volume two-body problem has been well-developed   \cite{Rummukainen:1995vs,Lin:2001ek,Christ:2005gi,Bernard:2007cm,Bernard:2008ax,He:2005ey,Lage:2009zv,Doring:2011vk,Aoki:2011gt,Briceno:2012yi,Hansen:2012tf,Guo:2012hv,Guo:2013vsa} based on a  pioneer work by L\"uscher \cite{Luscher:1990ux},  which is usually referred to  L\"uscher's formula. In fact, no matter what kind of boundary conditions are considered accordingly due to relativistic effect, moving frame, {\it etc.}, two-body L\"uscher's formula is the result of periodicity  properties of  asymptotic forms of wave function: (1) asymptotic form of wave function defines the physical transition amplitude and its general properties, such as unitarity relation, without  consideration of specific form of interaction; (2) the periodicity of wave function yields the secular equation that relates the scattering amplitudes to periodic lattice structure and produces   discrete energy spectra. As will be made clear later on, this   feature of finite volume dynamics is also true in multiple-body dynamics,  though because extra degrees of freedom and new types of interactions are introduced, the asymptotic form of multiple-body wave function appears more complex than two-body wave function. In general, the three-body problems are quite complex even in free space, the dynamics are usually described by off-shell unphysical amplitudes that are the solutions of Faddeev-type integral equations in momentum space. Finding the solutions of momentum space off-shell amplitudes in free space is already a uneasy task. Even in 1D space, only a few problems can be solved exactly, such as, McGuire's model in \cite{McGuire:1964zt,Guo:2016fgl}. On the other hand, the asymptotic form of wave function in configuration space is completely determined by physical transition amplitudes \cite{Gerjuoy:1971,Newton:1972xc}. Therefore, it seems natural to seek the solutions of finite volume three-body problems by considering asymptotic forms and periodicity of wave function in configuration space. The on-shell unitarity relation of physical amplitudes can be easily implemented by normalization of wave function in this way. The wave function approach has been proven valid and effective in finite volume two-body problems  \cite{Guo:2012hv,Guo:2013vsa}, and it has also been successfully employed to a solvable 1D three-body problem   in \cite{Guo:2016fgl}. Unfortunately, McGuire's model solved in \cite{Guo:2016fgl} yields no diffraction effect, no new momenta are generated, and wave function is simply given by sum of plane waves. This letter tends to show that  the wave function approach is originated from general features of multiple-body wave function, such as,  asymptotic behaviors and periodicity, therefore it must be valid  for finite volume multiple-body problems. Moreover, from mathematical perspective, the extra degrees of freedom in particle numbers is  equivalent to the two-body scattering in higher spatial dimensions with extra types of interaction. It hence may be   convenient to map the finite volume multiple-body problem to higher  spatial dimensions two-body problem. For a clear demonstration of wave function approach, we consider a one spatial dimensional three-body scattering of two spinless light particles and one infinite heavy particle, the interactions among three particles are given by two types: (1) the pair-wise interaction between one light and one heavy particle, denoted as $V$-potential; (2) a ``true'' three-body interaction that   all particles are involved in scattering, referred as $U$-potential.   This 1D three-body problem is  then mapped  into a 2D two-body scattering problem, pair-wise potential is associated to disconnected and rescattering contributions and yields plane waves type asymptotic behavior of wave function, and ``true'' three-body potential results in a diffractive spherical 2D scattered wave. The quantization of  three-body problem in finite volume is thus derived by taking into account of both asymptotic form  and periodicity of wave function in a 2D space.    In order to keep a clean and simple form in presentation, we have assumed that the pair-wise interaction between two light particles is absent, thus the disconnected contribution with heavy particle as a spectator and   rescattering contributions between  light-light pair and heavy-light pair vanish. The consideration of these contributions will only add some extra plane waves in asymptotic form of wave function, so neglect of these contributions won't affect the method that we tend to present,    the discussion with pair-wise interaction in all pairs will be given elsewhere.   In addition,  we will also ignore relativistic effects and work only in the center of mass frame of three-particle. As mentioned earlier, these   effects will only yield a  lattice with a distorted shape and make  a twist on periodic boundary condition, so that ignorance of these effects won't have much impact on our presentation as well.

{\it Three-body scattering  in free space.}---Considering scattering of two light and one infinite heavy spinless particles  in a 1D space, the heavy particle is    labeled as third particle, two distinguishable  light particles are labeled as  particle-1 and -2 with equal mass $m$.      The relative coordinates and momenta between light and heavy particles are denoted by $r_{1,2}$ and $q_{1,2}$ respectively.  The center of mass frame three-particle wave function satisfies Schr\"odinger  equation, 
\begin{equation}
 \left [  \frac{ \sigma^{2} + \nabla_{r_{1}}^{2} + \nabla_{r_{2}}^{2} }{2m}- \sum_{i=1}^{2}V(r_{i})  - U( \mathbf{ r} )  \right ] \psi (\mathbf{ r} ; \mathbf{ q}) =0 , \label{schrodiger}
\end{equation}
where \mbox{$\sigma^{2}=2 m E=q_{1}^{2}+q_{2}^{2}$} is associated  to the total  center of mass energy, and   short hand notation \mbox{$\mathbf{ r} =(r_{1},r_{2}) $} and \mbox{$ \mathbf{ q} = (q_{1}, q_{2})$} are adopted  throughout the presentation. Potential $V(r_{1,2})$ represents a pair-wise interaction between one light and the heavy particle,  and potential $U(\mathbf{ r})$ stands for a ``true'' three-body interaction with all three particles involved in scattering. Mathematically, the 1D three-body scattering problem given by Eq.(\ref{schrodiger}) is equivalent to a two-body scattering problem in a 2D space. The three-body wave function  is given by sum of multiple components, each component displays a distinct asymptotic form of scattered wave. As an example, for repulsive  interactions,     two distinct types of   scattered waves are (i) linear superposition of plane waves with no  new momenta created and describe scattering contribution due to pair-wise interactions; (ii)  the diffractive wave that resembles a spherical wave in 2D two-body scattering. Hence, the technique in 2D two-body scattering     can be  applied in 1D three-body scattering.

For   an incoming wave, $e^{i \mathbf{ q} \cdot \mathbf{ r}}$, the solution of three-body wave function is given by Lippmann-Schwinger equation,
\begin{align}
\psi (\mathbf{ r};\mathbf{ q}) & =  \phi_{q_{1}} (r_{1}) \phi_{q_{2}} (r_{2}) \nonumber \\
&+ \int d \mathbf{ r}'   G_{(12)} ( \mathbf{ r}, \mathbf{ r}' ;\sigma^{2})  2 m U (\mathbf{ r}') \psi(\mathbf{ r}'; \mathbf{ q}), \label{LippSchw}
\end{align}
where $ \phi_{q_{i}} (r_{i})$ stands for the wave function of two-body scattering between i-th \mbox{$(i=1,2)$}  and third particle, and it satisfies Schr\"odinger  equation \mbox{$\left ( q_{i}^{2} + \nabla_{r_{i}}^{2}   -2 m V(r_{i})  \right ) \phi_{q_{i}} (r_{i}) =0$}, and Green's function $G_{(12)}$ satisfies equation
\begin{align}
& \left [ \sigma^{2} + \nabla_{r_{1}}^{2} + \nabla_{r_{2}}^{2} -2 m V(r_{1}) -2 m  V(r_{2})  \right ] G_{(12)} ( \mathbf{ r}, \mathbf{ r}' ;\sigma^{2})   \nonumber \\
& \quad \quad \quad \quad \quad \quad = (2\pi) \delta(r_{1} - r'_{1}) (2\pi) \delta(r_{2} -r'_{2}).
\end{align}
Asymptotically, the solution for two-body wave function is   \cite{Guo:2013vsa,Guo:2016fgl}, $ \phi_{q_{i}} (r_{i})  \rightarrow e^{i q_{i } r_{i}} + i t( \frac{|q_{i}| r_{i}}{|r_{i}|}, q_{i}) e^{i |q_{i} r_{i}|} $,
where   the on-shell two-body scattering amplitude can be expanded in terms of parity eigenstates  \cite{Guo:2013vsa,Guo:2016fgl},  \mbox{$t(q'_{i}, q_{i})  =  \sum \limits_{\mathcal{P}_{i} = \pm} Y_{\mathcal{P}_{i}} (r_{i}) t_{\mathcal{P}_{i}} (|q_{i}|) Y_{\mathcal{P}_{i}} (q_{i})$},   where  \mbox{$Y_{+}(x)=1$} and \mbox{$Y_{-}(x)= \frac{x}{|x|}$}.

The first term on the right hand side of Eq.(\ref{LippSchw}) is composed of (1) an incoming wave, (2) the   disconnected scattering contribution with only one of light particles involved in scattering and another acting as a spectator, and (3) on-shell  three-body   rescattering  contribution when $U$-potential is completely turned  off, thus the three-body interaction   is realized by  rescattering effect with alternate scattering of one of two light particles off third particle. Rescattering effects are generated purely by pair-wise interactions and persist  even when $U$-potential is zero. The second term on the right hand side of Eq.(\ref{LippSchw}) represents   a ``true''  three-body interaction. Let's define a scattering $T$-amplitude that is associated to $U$-potential by 
\begin{align}
&- \frac{T_{(12)} ( \mathbf{ k}; \mathbf{ q}) }{\sigma^{2}-k^{2} }  \nonumber \\
& \quad   =   \int d \mathbf{ r}   d \mathbf{ r}'  e^{- i \mathbf{ k} \cdot \mathbf{ r}}  G_{(12)} ( \mathbf{ r}, \mathbf{ r}' ;\sigma^{2})  2 m U (\mathbf{ r}') \psi(\mathbf{ r}'; \mathbf{ q})   ,
\end{align}
where $k^{2}=k_{1}^{2}+k_{2}^{2}$. Hence, we can rewrite  Eq.(\ref{schrodiger}) to
\begin{equation}
\psi (\mathbf{ r};\mathbf{ q}) =  \phi_{q_{1}} (r_{1}) \phi_{q_{2}} (r_{2}) -   \int \frac{ d \mathbf{ k} }{(2\pi)^{2}} e^{ i \mathbf{ k} \cdot   \mathbf{ r}  }  \frac{ T_{(12)} ( \mathbf{ k}; \mathbf{ q})}{\sigma^{2} - k^{2}   } .  \label{waveT12}
\end{equation}
It can be easily show that \cite{Gerjuoy:1971,Newton:1972xc} 
\begin{equation}
 T_{(12)} ( \mathbf{ k}; \mathbf{ q})  =-  \int d \mathbf{ r}'    \phi_{ -  k_{1}}(r'_{1}) \phi_{-k_{2}} (r'_{2})  2 m U (\mathbf{ r}') \psi(\mathbf{ r}'; \mathbf{ q}). 
\end{equation}

For repulsive  interactions,  because of the absence of two-body bound states, $T_{(12)} $ displays no primary  singularities \cite{Faddeev:1960su,Faddeev:1965}, and describes the   ``true'' three-to-three particles scattering process, which will be denoted by \mbox{$T_{0,0}  $} from now on. The   asymptotic form of wave function thus is given by, 
\begin{align}
\psi (\mathbf{ r};\mathbf{ q})    \rightarrow   \phi_{q_{1}} (r_{1}) \phi_{q_{2}} (r_{2})  + \frac{i e^{i (\sigma r - \frac{\pi}{4})}}{2\sqrt{2\pi \sigma r}}  T_{0,0} (     \frac{\sigma \mathbf{ r}}{r} ; \mathbf{ q}),
\end{align}
 where \mbox{$r=\sqrt{r_{1}^{2} + r_{2}^{2}}$}. The $T_{0,0}$-amplitude is constrained by unitarity relation,
 \begin{align}
 & \sum_{  k_{1,2} =\pm |q_{1,2}| }    \left [ s_{V}^{\dag}( \mathbf{ k}; \mathbf{ q})   T_{0,0}( \mathbf{ k}; \mathbf{ q}' ) -     T^{\dag}_{0,0}( \mathbf{ k}; \mathbf{ q} ) s_{V} ( \mathbf{ k}; \mathbf{ q}')   \right ]\nonumber \\
 &  \quad \quad \quad  \quad    =  \frac{ i}{2} \int_{0}^{2\pi} \frac{ d \theta_{k}}{2\pi}   T_{0,0}^{\dag}( \mathbf{ k}; \mathbf{ q})   T_{0,0}( \mathbf{ k}; \mathbf{ q}')  ,
\end{align}
where the reduced  two-body $S$-matrix, $s_{V}$, is  
\begin{align}
s_{V}( \mathbf{ q}'; \mathbf{ q})  &= \delta_{  q'_{1}   ,  q_{1} }  \delta_{  q'_{2}   ,  q_{2} } +\delta_{  q'_{1}   ,  q_{1} } i t ( q'_{2}, q_{2} ) +  \delta_{  q'_{2}   ,  q_{2} }   i  t ( q'_{1}, q_{1} )   \nonumber \\
& +  i  t ( q'_{1}, q_{1} )  i t ( q'_{2}, q_{2} )  ,
\end{align}
the Kronecker delta $\delta_{q'_{i}, q_{i}}$ is equal to one if $q'_{i}$ and $q_{i}$ have the same sign, and  zero otherwise, and 
\begin{align}
 \sum \limits_{ k_{1,2} =\pm |q_{1,2}| }  s_{V}^{\dag}( \mathbf{ k}; \mathbf{ q})  s_{V}( \mathbf{ k}; \mathbf{ q}' )   = \delta_{  q'_{1}   ,  q_{1} }  \delta_{  q'_{2}   ,  q_{2} }.
 \end{align}
    In this letter, the three-body problem with  only  repulsive interactions will be considered.  The treatment and  formalism for attractive potentials will be presented  in \cite{Guo:2017}.   The first  term in, $\phi_{q_{1}} (r_{1}) \phi_{q_{2}} (r_{2})$, appears as the product of two two-body wave functions,  hence in finite volume,  it can be handled rather easily with technique that have been developed for two-body scattering, see  \cite{Guo:2012hv,Guo:2013vsa,Guo:2016fgl}. 
However, the diffractive  wave, the  ``true'' three-body scattering contribution,   behaves  as a 2D spherical wave and thus has to be treated differently. Taking advantage of resemblance of diffractive wave of 1D three-body scattering  and spherical wave of 2D two-body scattering,   the diffractive term of 1D three-body scattering will be treated    just as a 2D two-body scattering problem mathematically.   Therefore, similar to partial wave expansion in 2D two-body  scattering, see Appendix,     a ``partial wave expansion'' of on-shell $T_{0,0}$ amplitude  is  introduced by,
\begin{align}
& T_{0,0} ( \frac{\sigma \mathbf{ r}}{r} ; \mathbf{ q})  = 4  \sum_{ J' = - \infty}^{\infty}  e^{i J' \theta_{r}  } T_{J'}(\mathbf{ q})   ,  \  \tan \theta_{r} =  \frac{r_{2}}{r_{1}}, \nonumber \\
& T_{J'}(\mathbf{ q})    =  \sum \limits_{ J= - \infty}^{\infty}    T_{J', J}(\sigma)  e^{- i J \theta_{q}}, \ \tan \theta_{q} =   \frac{q_{2}}{q_{1}} .  \label{pwT00} 
\end{align}
The three-body wave function for repulsive  interactions is thus given by
\begin{equation}
\psi (\mathbf{ r};\mathbf{ q})      =  \phi_{q_{1}} (r_{1})   \phi_{q_{2}} (r_{2})+    \sum_{J = -\infty}^{\infty}    i^{J} e^{i J  \theta_{r}}    H_{J}^{(1)} (\sigma  r )    i  T_{J}(\mathbf{ q})  . \label{freewav}
\end{equation}

 {\it Three-body scattering in finite volume.}---For   three-particle interaction in a periodic box with   size, $L$,  the  three-particle  wave function in finite volume,   $\psi^{(  L)}  $, must satisfy periodic boundary condition, \mbox{$\psi^{( L)} (\mathbf{ r} + \mathbf{ n} L;\mathbf{ q}) = \psi^{( L)} (\mathbf{ r};\mathbf{ q})$}, and $\psi^{( L)}  $ can be constructed by free space three-body  wave function, \mbox{$\psi^{( L)} (\mathbf{ r};\mathbf{ q})    = \sum \limits_{\mathbf{ n} \in \mathbb{Z}} \psi  (\mathbf{ r} + \mathbf{ n} L ;\mathbf{ q})$}, thus
\begin{align}
 & \psi^{( L)} (\mathbf{ r};\mathbf{ q})    
=    \phi^{(L)}_{q_{1}} (r_{1}  )      \phi^{(L)}_{q_{2}} (r_{2}  ) \nonumber \\
& +  \frac{i}{4} \int \frac{ d \mathbf{ r}'   d \mathbf{ k} }{(2\pi)^{2}}  \sum_{\mathbf{ n} \in \mathbb{Z}} H_{0}^{(1)} ( \sigma  | \mathbf{ r}  + \mathbf{ n} L - \mathbf{ r}' |)   e^{ i \mathbf{ k} \cdot \mathbf{ r}'} T_{0,0} ( \mathbf{ k}; \mathbf{ q}),
\end{align}
where the  finite volume  two-body wave function  is given analytically  \cite{Guo:2012hv,Guo:2013vsa,Guo:2016fgl} by 
\begin{align}
    \phi^{(L)}_{q_{i}} (r_{i})  =  & \sum_{ n_{i}  \in \mathbb{Z}}   \phi_{q_{i}} (r_{i} + n_{i} L)  =  it  ( q'_{i}, q_{i} )    e^{i | q_{i} r_{i}   |}    \nonumber \\
   &  +   \sum_{\mathcal{P}_{i}=\pm}    it_{\mathcal{P}_{i}} ( | q_{i}| )    \frac{e^{i q_{i}  r_{i}  }  + (-1)^{ \mathcal{P}_{i}}  e^{-i q_{i} r_{i}   }   }{e^{- i |q_{i}| L }-1}  .
\end{align}
 The infinite sum in second term is carried out by using Eq.(\ref{expancoef}),  also with the help of   ``partial wave expansion''  of $T_{0,0} $ in Eq.(\ref{pwT00}), we obtain  
\begin{align}
& \psi^{(  L)} (\mathbf{ r};\mathbf{ q}) = \phi^{(L)}_{q_{1}} (r_{1})   \phi^{(L)}_{q_{2}} (r_{2})        \nonumber \\
& +  \!\!\!\!  \sum_{J , J' = - \infty}^{\infty} \!\!\!\!   e^{i J  \theta_{r}   }  \left [ \delta_{J, J'} i N_{J} ( \sigma  r) +   g_{J'-J}   ( \sigma  )  J_{J}  (\sigma r)     \right ]     i^{J'}  i T_{J'}( \mathbf{ q})     . \label{ftwav}
\end{align}

The quantization condition of three-body scattering is obtained by matching $\psi^{(  L)} $ to free space three-body wave function given   in Eq.(\ref{freewav}), \mbox{$\psi^{(  L)} (\mathbf{ r};\mathbf{ q}) = \psi (\mathbf{ r};\mathbf{ q}) $}.  After the ``partial wave'' projection, the matching condition has   non-trivial solutions for an arbitrary $r$ only when
 \begin{align}
     &   \det \Bigg\{  \delta_{J, J'}    \left ( e^{- i J \theta_{q}}+ i T_{J}( \mathbf{ q})      + \sum_{j  = - \infty}^{\infty}  i^{j-J}    g_{j-J}   ( \sigma  )   i T_{j}( \mathbf{ q})      \right )      \nonumber \\
&    \quad\quad\quad\quad\quad \quad  \ +     \sum_{\mathcal{P}_{1} , \mathcal{P}_{2} = \pm }    \mathcal{M}^{( \mathcal{P}_{1},\mathcal{P}_{2} )}_{J, J'}  (\sigma, \theta_{q})  \Bigg \}   =  0   , \label{matchcond}
\end{align}
where  $T_{L}$'s are associated to the diffractive wave contribution and describes the ``true'' three-body scattering, and $  \mathcal{M}^{( \mathcal{P}_{1},\mathcal{P}_{2} )}_{J, J'}  $   is associated to the disconnected  and rescattering  contributions that are determined solely by pair-wise two-body interactions, 
\begin{align}
& \mathcal{M}^{(\mathcal{P}_{1},\mathcal{P}_{2} )}_{J, J'}  (\sigma, \theta_{q})   = -  \delta_{J, J'}     \frac{2   i t_{\mathcal{P}_{1} }(| q_{1}|)      }{e^{- i |q_{1}| L }-1}    \frac{2   i t_{\mathcal{P}_{2} }(| q_{2}|)      }{e^{- i |q_{2}| L }-1}   \nonumber \\
& \quad   \quad \quad  \times  \frac{  \frac{1+(-1)^{\mathcal{P}_{1}  + \mathcal{P}_{2} +J} }{2}   e^{- i J \theta_{q}}  +    \frac{ (-1)^{\mathcal{P}_{2} }  + (-1)^{\mathcal{P}_{1}+ J} }{2} e^{ i J \theta_{q}} }{2}   \nonumber \\
& +        i t_{\mathcal{P}_{1} }( | q_{1}|)    \left (1   -   \frac{2   i t_{\mathcal{P}_{2} }( |q_{2} | )     }{e^{- i |q_{2}| L }-1}  \right )  A_{J, J'}^{( \mathcal{P}_{1} ,\mathcal{P}_{2} )} ( \theta_{q} )   \nonumber \\
 & +        i t_{\mathcal{P}_{2} }( | q_{2}|)     \left (  1- \frac{2   i t_{\mathcal{P}_{1} }(| q_{1}|)      }{e^{- i |q_{1}| L }-1} \right )  B_{J, J'}^{( \mathcal{P}_{1} ,\mathcal{P}_{2} )} (   \theta_{q} )  ,
\end{align}
and ``partial wave expansion''  coefficients, $ A_{J, J'}^{( \mathcal{P}_{1},\mathcal{P}_{2}  )} $ and $ B_{J, J'}^{( \mathcal{P}_{1} ,\mathcal{P}_{2} )} $, are defined by relations,
\begin{align}
&  Y_{\mathcal{P}_{1} } (r_{1}) Y_{\mathcal{P}_{1} } (q_{1}) e^{  i | q_{1 } r_{1}|}  \frac{ e^{  i q_{2} r_{2}} + (-1)^{\mathcal{P}_{2} } e^{ - i q_{2} r_{2} } }{2} \nonumber \\
& \quad \quad  \quad \quad \quad     =  \sum_{J, J' = - \infty}^{\infty}  i^{J}e^{i J  \theta_{r}   }   A_{J, J'}^{( \mathcal{P}_{1} ,\mathcal{P}_{2} )} (  \theta_{q} )  J_{J'}  (\sigma r) ,   
\end{align}
\begin{align}
&Y_{\mathcal{P}_{2} } (r_{2}) Y_{\mathcal{P}_{2} } (q_{2}) e^{  i | q_{2 } r_{2}|}  \frac{ e^{  i q_{1} r_{1}} + (-1)^{\mathcal{P}_{1} } e^{ - i q_{1} r_{1} } }{2}     \nonumber \\
&\quad \quad  \quad \quad \quad   = \sum_{J, J' = -\infty}^{\infty}  i^{J}e^{i J  \theta_{r}   }   B_{J, J'}^{( \mathcal{P}_{1} ,\mathcal{P}_{2} )} (  \theta_{q} )  J_{J'}  (\sigma r) .
\end{align}

 {\it Discussion and conclusion}---As  already been discussed in  \cite{Guo:2012hv,Guo:2013vsa,Guo:2016fgl},  L\"uscher's formula in two-body scattering is the consequence of general feature of periodicity and asymptotic form of wave function, which doesn't depend on a specific form of interaction. This feature is in fact still valid for finite volume three-body scattering, however, due to   extra degrees of freedom, the asymptotic form of three-body wave function display  more complex structures.  Fortunately, from mathematics point of view, these extra degrees of freedom in particle numbers can be dealt  equivalently as two-body scattering in higher dimensions.  Although, from physical perspective,   interaction among three particles are usually   more sophisticated   than the interaction  considered   in  two-body scattering, the difference in the form of interactions results in the distinct   asymptotic form of wave functions that correspond to different physical processes.       

For a more explicit demonstration, let's consider two extreme limits of our three-body model:  

(1) \mbox{$U  = 0$: } 
The solution of wave function is given by $\phi_{q_{1}}(r_{1}) \phi_{q_{2}}(r_{2})$.    In addition to a free incoming wave, the three-body wave function consists of two types of contribution, one is   disconnected contributions with one of two light particles acting as a spectator, \mbox{$i t (q'_{1}, q_{1}) e^{i |q_{1} r_{1}|} e^{ i q_{2} r_{2}}$} and  \mbox{$i t (q'_{2}, q_{2}) e^{i |q_{2} r_{2}|} e^{ i q_{1} r_{1}} $}. The second type is three-body rescattering contribution with   alternate scattering of one   light particle off third one, so   all three particles are involved in scattering by iterations, \mbox{$i t (q'_{1}, q_{1}) i t (q'_{2}, q_{2})  e^{i |q_{1} r_{1}|}   e^{i |q_{2} r_{2}|}   $}.   As the consequence of  no direct interaction between two light particles, no new momenta are created after  collision,     rearrangement of momenta between two light particles are   not allowed,  and diffraction effects vanish in this particular case.   The determinant condition in Eq.(\ref{matchcond}) reduce to 1D L\"uscher's formula like quantization condition   \cite{Guo:2013vsa,Guo:2016fgl}, $e^{- i |q_{1,2}| L} = 1+ 2i t_{\pm} (|q_{1,2}|)$, and the momentum of each light particle  appears discrete, so is total energy of three-particle.

(2) \mbox{$V = 0$:}   
 As the result of no pair-wise interactions at all, all the   two-body scattering $t(q'_{i}, q_{i})$-amplitude vanish, so that both disconnected and rescattering contribution are gone, and \mbox{$\phi_{q_{1}} (r_{1})\phi_{q_{2}} (r_{2}) = e^{i \mathbf{ q} \cdot \mathbf{ r}}$}. In this case, the 1D three-body wave function in Eq.(\ref{freewav}) indeed resembles  two-body wave function  in 2D space presented in Appendix. Mathematically, the problem of three-body scattering in finite volume thus can be solved   as a problem of  two-body scattering in a 2D space. The matching condition in Eq.(\ref{matchcond}) thus reduces to a form that resembles to matching condition in 2D scattering,  
\begin{align}
     &   \det    \left ( \delta_{J, J'}+ i T_{J, J'}(\sigma)   -   \sum_{j  = - \infty}^{\infty}  i^{j-J}    g_{j-J}   ( \sigma  )   i T_{j, J'}( \sigma)        \right )         =  0   . 
\end{align}
If $U$-potential is spherical, \mbox{$U(\mathbf{ r}) = U(r)$},  hence, $T_{J, J'}=\delta_{J,J'} T_{ J'}$,  and above 1D three-body quantization condition is thus equivalent to Eq.(\ref{lushcer2D}),   and the unitarity relation for partial wave scattering $T_{J}$-matrix is given also in a simple form,  \mbox{$ \mbox{Im} T_{J} =  T^{*}_{J} T_{J} $} and \mbox{$T_{J} = \frac{1}{\cot \delta_{J}-i}$}.

In general, when   $V$-and $U$-potential are both non-zero, finding solutions of the quantization condition Eq.(\ref{matchcond}) is    not a easy task,  even though  the   model that is considered in present work is already quite simple. Nevertheless, the  generalization of this simple model to 3D and also including spin of particles may still be a good physical description for certain processess of three-body interaction, such as, the $Y(4260)$ production in $J/\Psi \pi \pi$ three-body state, in which the  $Y(4260)$   may be described by a $U$-type three-body  interaction,   interactions between $J/\Psi$ and one of $ \pi$'s  can be modeled by $V$-type two-body interaction that is responsible for the production of $Z_{c}(3900)$ state in  $J/\Psi \pi $ sub-channel. As a good approximation, see Fig.2 in \cite{Ablikim:2013mio}, $\pi \pi$ interaction can be   ignored  in $Y(4260) \rightarrow J/\Psi \pi \pi$ process. We leave  the further discussion of the strategies of finding solutions of  the quantization condition Eq.(\ref{matchcond})  to \cite{Guo:2017}.  Until then, let's consider a special situation so we may get a sense of what we may expect in general.  Assuming that rescattering effect is teated as perturbation for weak $V$-potential, and with further assumption of the spherical symmetric $U$-potential for scattering of three bosons, thus, the dominant   contributions are from diagonal terms in Eq.(\ref{matchcond}). Therefore, we end up with a expression,
 \begin{align}
       \det \bigg \{   \sum_{j  = - \infty}^{\infty} &    \left [ \delta_{j,J}  +\delta_{j,J}    i T_{j}( \sigma)                 +   i^{j-J}    g_{j-J}   ( \sigma  )   i T_{j}( \sigma)\right ]     \cos j \theta_{q}  \nonumber \\
&  \  +  \mathcal{M} (\sigma, \theta_{q})     \cos J \theta_{q}    \bigg \} =0  . \label{disc}
\end{align}
where
\begin{align}
 \mathcal{M}  (\sigma, \theta_{q})   & =  -    \frac{2   i t_{+ }(| q_{1}|)      }{e^{- i |q_{1}| L }-1}    \frac{2   i t_{+}(| q_{2}|)      }{e^{- i |q_{2}| L }-1}      \nonumber \\
& +       i t_{+}( | q_{1}|)    \left (1   -   \frac{2   i t_{+ }( |q_{2} | )     }{e^{- i |q_{2}| L }-1}  \right )      \nonumber \\
& +         i t_{+ }( | q_{2}|)     \left (  1- \frac{2   i t_{+ }(| q_{1}|)      }{e^{- i |q_{1}| L }-1} \right )    . \label{diagonalM}
\end{align}
Eq.(\ref{disc}-\ref{diagonalM})  provide a rough idea how the solution of $U$-potential may be shifted by  weak two-body interaction of $V$-type potential in perturbation. As we can see clearly from Eq.(\ref{disc}),  the solutions in previously discussed  two extreme limits of either \mbox{$U=0$} or \mbox{$V=0$} do not satisfy   quantization conditions in general. The assumption of weak potentials and perturbation expansion may be justified by physical processes in condensed matter, such as,  photonic crystals with Rashba-like spin-orbit interaction term \cite{Joannopoulos:2008,Manchon:2015}.  Even when the interaction is weak, it still introduces a small correction to the band structure of the two-dimensional Bloch states (splitting of spin bands) or creates the dynamical phase shift between the waves propagating in the orthogonal directions. The investigations in these physical processes are stimulated by the possibility of creation of spin/polarization controlling devices, where the electronÕs spin/polarization of the propagating light could be precisely manipulated and controlled.

 In summary,  we propose the wave function approach to the solutions of  finite volume three-body problem, this approach is based on general properties of wave function in configuration space, such as, asymptotic forms and periodicity that are related to  on-shell physical transition amplitudes  and periodic lattice structure respectively. From mathematical perspective, multiple-body problem is equivalent to a two-body problem in higher dimension, so it is convenient to map a three-body problem into a higher dimensional two-body problem. The solutions of finite volume three-body problem derived by wave function approach require only on-shell physical amplitudes and resemble two-body  L\"uscher's formula in higher dimensions.   The idea is demonstrated by using a simple 1D three-body problem with interaction of  two light spinless particles and one heavy particle  as a example.

{\it ACKNOWLEDGMENTS}---We   acknowledges support from Department of Physics and Engineering, California State University, Bakersfield, CA.

{\it Appendix: Finite volume two-body scattering in two spatial dimensions.}---Let's consider   two spinless particles scattering in a 2D space, the interaction between two particles is described by spherical $U$-potential.    The     two-body wave function in center of mass  frame satisfies
 \begin{align}
& \psi (\mathbf{ r};\mathbf{ q})  =  e^{i \mathbf{ q} \cdot \mathbf{ r}}-   \int \frac{ d \mathbf{ k} }{(2\pi)^{2}}   \frac{ e^{ i \mathbf{ k} \cdot   \mathbf{ r}  } }{\sigma^{2} - k^{2}   } T ( \mathbf{ k}; \mathbf{ q}) ,  \label{2Dwave}
\end{align}
where $\mathbf{ r}$ and $\mathbf{ q}$ refer to the relative coordinate and  momenta of two particles respectively, and \mbox{$\sigma^{2}=2 m E= q^{2}$}.  The two-body scattering amplitude is defined by \mbox{$ T  ( \mathbf{ k}; \mathbf{ q})  =-  \int d \mathbf{ r}   e^{- i \mathbf{ k} \cdot \mathbf{ r}}  2 m U (r) \psi(\mathbf{ r}; \mathbf{ q})$}.  The partial wave expansion of   two-body wave function in 2D  reads
\begin{align}
& \psi  (\mathbf{ r};\mathbf{ q})   =     \sum_{J= -\infty}^{\infty}  i^{J} e^{i J  \theta_{r}}   \left [ J_{J} (\sigma  r )  +    i  T_{ J}(\sigma)   H_{J}^{(1)} (\sigma  r ) \right ] e^{- i J \theta_{q}} ,
\end{align}
where    $T_{J}$-amplitude can be parameterized by phase shifts, \mbox{$T_{J} = \frac{e^{2 i \delta_{J}}-1}{2i}$}, and is related to $T$-amplitude by  \mbox{$ T  ( \mathbf{ q}'; \mathbf{ q})  = 4\sum \limits_{J = - \infty}^{\infty}  e^{i J \theta_{q'}  } T_{J}(\sigma) e^{- i J \theta_{q}}$}.

When the particles are placed in a 2D periodic box of   size, $L$,  the finite volume two-particle wave function,  $ \psi^{(L)}  (\mathbf{ r};\mathbf{ q}) $, thus satisfies periodic boundary condition, \mbox{$\psi^{(L)}  (\mathbf{ r} + \mathbf{ n} L;\mathbf{ q}) = \psi^{(L)}  (\mathbf{ r};\mathbf{ q})$}  \mbox{$(\mathbf{ n} \in \mathbb{Z})$}. The finite volume two-body wave function can be constructed from free space wave function, \mbox{$\psi^{(L)}  (\mathbf{ r};\mathbf{ q})  = \sum \limits_{\mathbf{ n} \in \mathbb{Z}} \psi  (\mathbf{ r} + \mathbf{ n} L;\mathbf{ q})$}, thus
\begin{align}
\psi^{(L)}  (\mathbf{ r};\mathbf{ q})     = &        \frac{i}{4} \int d \mathbf{ r}'    \sum_{\mathbf{ n} \in \mathbb{Z}}   H_{0}^{(1)} ( \sigma  | \mathbf{ r}  + \mathbf{ n} L- \mathbf{ r}' |)  \nonumber \\
& \times   \int \frac{ d \mathbf{ k} }{(2\pi)^{2}} e^{ i \mathbf{ k} \cdot \mathbf{ r}'}  T ( \mathbf{ k}; \mathbf{ q}) .
\end{align}
The infinite sum is carried out by partial wave expansion as in 3D \cite{Luscher:1990ux,Guo:2012hv},
\begin{align}
&  \frac{4 i }{L^{2}}  \sum_{\mathbf{ k} = \frac{2\pi}{L} \mathbf{ n}}^{\mathbf{ n} \in \mathbb{Z}} \frac{ e^{ i \mathbf{ k} \cdot ( \mathbf{ r} -\mathbf{ r}' )  }   }{\sigma^{2} - k^{2}}  = \sum_{\mathbf{ n} \in \mathbb{Z}}   H_{0}^{(1)} ( \sigma  | \mathbf{ r}  + \mathbf{ n} L- \mathbf{ r}' |)    \nonumber \\
& = \sum_{J , J' = - \infty}^{\infty}   e^{i J  \theta_{r}   }  \left [ \delta_{J, J'} i N_{J} ( \sigma  r) +   g_{J'-J}   ( \sigma  )  J_{J}  (\sigma r)     \right ]    \nonumber \\
& \quad \quad \times  J_{J'}( \sigma  r') e^{ - i J'   \theta_{r'} }, \label{expancoef}
\end{align}
where the expansion coefficient, $g_{J}$, is given by
 \begin{align}
&    g_{J}   (\sigma)  =      \frac{4 i }{L^{2}} \sum_{\mathbf{ k} =\frac{2 \pi}{L}  \mathbf{ n}}^{\mathbf{ n} \in \mathbb{Z} }   \frac{i^{J} ( \frac{\sqrt{k^{2}}}{ \sigma } )^{J}  e^{ - i J   \theta_{k}} }{\sigma^{2} - k^{2}  }  -  \delta_{J,0}  i  N_{0} (\sigma r ) |_{r\rightarrow 0} .
\end{align}
 Therefore, we obtain the partial wave expansion of finite volume two-body wave function in 2D,
\begin{align}
 \psi^{(L)}  (\mathbf{ r};\mathbf{ q})    
&= \!\!\!\! \sum_{J , J' = - \infty}^{\infty} \!\!\!\!   e^{i J  \theta_{r}   }  \left [ \delta_{J, J'} i N_{J} ( \sigma  r) +   g_{J'-J}   ( \sigma  )  J_{J}  (\sigma r)     \right ]  \nonumber \\
&  \quad   \times  i^{J'}     i  T_{ J'}(\sigma)     e^{- i J' \theta_{q}} .
\end{align}

The secular equation is obtained by matching $\psi^{(L)}  (\mathbf{ r};\mathbf{ q})   $ to $\psi   (\mathbf{ r};\mathbf{ q})   $ at an arbitrary $\mathbf{ r}$, thus we get
\begin{align}
\det \left [ \delta_{J, J'}  \cot \delta_{J} (\sigma) -   i g_{J'-J}   ( \sigma  )     \right ]  =0. \label{lushcer2D}
\end{align}
 Eq.(\ref{lushcer2D}) is a 2D version of  L\"uscher's formula like quantization condition for two spinless particles scattering.

\end{document}